\documentclass[aps,pra,twocolumn,floatfix,showpacs,superscriptaddress,unsortedaddress]{revtex4}
\usepackage{graphicx}
\usepackage{psfrag}
\usepackage{amsmath}

\newcommand{\prL}{Phys.\ Rev.\ Lett.\ }
\newcommand{\pr}{Phys.\ Rev.\ }
\newcommand{\jpb}{J.\ Phys.\ B: Atom.\ Mol.\ Opt.\ Phys.}
\newcommand{\epl}{Europhys.\ Lett.\ }

\newcommand{\phr}{Phys.\ Rep.\ }

\newcommand{\anp}{Ann.\ Phys.\ }

\newcommand{\zpd}{Z.\ Phys.\ D\ }

\newcommand{\jpa}{J.\ Phys.\ A\ }

\newcommand{\phd}{Physica D\ }

\begin{document}

\title{A realistic example of chaotic tunneling:\\ The hydrogen atom in
parallel static electric and magnetic fields}

\author{Dominique Delande}
\affiliation{Laboratoire Kastler-Brossel,
Tour 12, Etage 1, 4 Place Jussieu, F-75005 Paris, France}
\author{Jakub Zakrzewski}
\affiliation{Instytut Fizyki imienia Mariana Smoluchowskiego, Uniwersytet
 Jagiello\'nski, ulica Reymonta 4, PL-30-059 Krak\'ow, Poland}

\date{\today}

\begin{abstract}
Statistics of tunneling rates in the presence of chaotic classical
dynamics is discussed on a realistic example: a hydrogen atom
placed in parallel uniform static electric and magnetic fields, where
tunneling is followed by ionization along the fields direction.
Depending on the magnetic quantum number, one may observe either
a standard Porter-Thomas distribution of tunneling rates or,
for strong scarring by a periodic orbit parallel to the external
fields, strong deviations from it. For the latter case, a simple model based on
random matrix theory gives the correct distribution.
\end{abstract}


\maketitle 
\section{Introduction}

After many years of intensive research in the ``quantum chaos area''
it is now commonly accepted
that the quantum behavior of complex systems may be strongly correlated
with the character of their classical motion
 \cite{bohigas91,haake90,gaspard98,brack97,stoeckmann99}.
  Even such a purely quantum
phenomenon as tunneling may be profoundly affected by chaotic
classical dynamics. For regular systems
a smooth dependence of the tunneling rate
on parameters is expected. In the presence of chaotic motion, the
tunneling rates typically strongly fluctuate, the game is then to
identify both the average behavior and the statistical properties of the fluctuations.

Imagine the situation when the wavefunction is predominantly
localized in a region of regular motion. The tunneling to the chaotic
sea surrounding the regular island, called ``chaos assisted tunneling''
(CAT) has been quite thoroughly studied
\cite{bohigas90a,bohigas93,tomsovic94,averbukh95,leyvraz96,kuba98,kuba98a,frischat98,dembowski00,mouchet03}.
It may be characterized by the statistics of tunneling rates, or
directly measurable quantities such as tunneling splittings between
doublets of different symmetries \cite{leyvraz96} or tunneling widths
\cite{kuba98,kuba98a} where the tunneling to the chaotic sea leads eventually
to decay (e.g. to ionization of atomic species). Model based on random matrix theory (RMT)
\cite{brody81,mehta91} show that distributions of both quantities are
closely correlated with both the splittings \cite{leyvraz96} and {\it
square roots} of the widths \cite{kuba98}
having a common Cauchy (lorentzian-like) 
distribution with an exponential cutoff for extremely large events.
Such a situation occurs for sufficiently small $\hbar$ (in the semiclassical
regime) when the tunneling coupling is much smaller than the mean level
spacing in a given system.

Another possibility occurs when virtually all accessible phase space (at a
given energy) is chaotic: the tunneling occurs through a potential (rather
than dynamical as in the previous case) barrier. Then a standard RMT based
answer leads to the Porter-Thomas distribution of widths (or its appropriate
many channel extension) as applied in various areas from
nuclear physics \cite{brody81}, mesoscopics  \cite{jalabert92}
or chemical reactions \cite{nakamura00} to name a few. 
Creagh and Whelan \cite{creagh96a,creagh99a,creagh99b,creagh02} developed
a semiclassical approach to tunneling (for a semiclassical
 treatment concentrating on other aspects of tunneling
 see e.g. \cite{shudo95,shudo98}) which enabled them to give
 an improved statistical distribution of tunneling rates~\cite{creagh00a}. The distribution
 has been tested on a model system and shown to faithfully represent the
 tunneling splitting distribution  provided the classical dynamics is
 sufficiently chaotic. However, this distribution fails for systems
 when scarred \cite{heller84,bogomolny88a,kaplan98a,kaplan98b,bies01} wavefunctions
 dominate the process. In order to take into account scarring,
the same authors~\cite{creagh02} developed a more complicated semiclassical
theory which, in a model system, accurately describes the numerically observed  
tunneling rates.

 The   aim of this paper is twofold. Firstly, we propose a simpler  approach
 to
 the effect of scarring than 
 that
 in~\cite{creagh02}. Our approach is less general,
 as it is limited to the case when only one channel contributes to tunneling.
 This is, however, a very frequent situation: because tunneling typically
 decays exponentially with some parameter, most contributions are often
 hidden by a single dominant one. The formulas that we obtain are also
 much simpler. Secondly, we consider the tunneling rate distribution
 in a challenging, realistic system - a hydrogen atom in parallel electric
 and magnetic fields. As mentioned by Creagh and Whelan, one expects there
 the above mentioned problems due to scar-dominated tunneling. Here again
 we test the proposed distribution on
 a vast set of numerical data. Thirdly, in contrast with most of
 the previous studies,
 we do not consider here a situation where tunneling manifests itself
  as a quasi-degeneracy
 between a pair of even-odd states, but 
  rather the case when
  tunneling is followed by a subsequent
 ionization of the system and manifests itself in the widths (ionization rates)
 of resonances. The analysis for both cases is similar, but not identical.

 \section{The distribution for tunneling rates for scar dominated chaotic
 tunneling}

 Let us recall first shortly the analysis of chaotic
 tunneling used in~\cite{creagh00a}, which makes it possible to predict
 the distribution of tunneling rates
 in terms of classical quantities. This approach is based on 
 the standard semiclassical expansion of the Green function
as a sum over classical orbits (which is used e.g. in periodic orbit theory
{\it \`a la Gutzwiller}), but incorporates in addition some complex orbits,
that is orbits where time, position and momentum can be made complex.
Such orbits may tunnel through the potential well and eventually lead to
escape at infinity; they are essential for the calculation of tunneling rates.
In the one-dimensional case, it is well understood that tunneling can be quantitatively
described using a single complex orbit known as the instanton: the orbit propagates
under the potential well with a purely real position, and purely imaginary time and momentum,
until it emerges in the real phase space when the potential barrier is crossed
(it can be visualized as a standard real orbit in the inverted potential).
The action $S_I$ of the instanton is then purely imaginary $S_I=iK$ and the
tunneling rate is, not surprisingly, 
essentially described by the $\exp(iS_I/\hbar)=\exp(-K/\hbar)$
contribution. 
 
For a multidimensional system, the situation is somehow comparable,
except that there are now several instanton orbits. It also turns out that the
structure of the tunneling complex orbits can be extremely complicated~\cite{shudo98,brodier02}. 
However, because of the exponential
decrease of the tunneling rate, in the semiclassical
limit $\hbar \to 0,$ there are cases when the instanton orbit with the smallest imaginary action will give the 
dominant contribution. Creagh and Whelan succeeded in expressing the tunneling 
rate in terms of the action and stability
 exponent of the instanton orbit~\cite{creagh99a}. They were able
 to describe the situation of a symmetric double well, where tunneling manifests 
 itself through the existence of pairs of quasi-degenerate states, i.e. to calculate
 the splitting of the doublets. Comparison with ``exact" numerical results for a model
 system showed a very good agreement~\cite{creagh96a,creagh99a}. They were also able
 to describe the situation of tunneling outside a single potential well (with
 chaotic dynamics inside the well) followed by ``ionization", that is particle directly escaping
 toward infinity. The quantity of interest is the ``weighted" density of states,
 where the weight is given by the widths $\Gamma_n$ of the resonances with energies
 $E_n$:
\begin{equation}
 f(E)=\sum_n \Gamma_n \delta(E-E_n)
 \label{cw1}
 \end{equation}
 In the semiclassical approximation, it can be written --
  in the spirit of periodic orbit theory -- as the sum of  smooth 
 and oscillatory terms:
 \begin{equation}
 f(E) \approx f_0(E)+f_{\mathrm{osc}}(E),
 \label{cw2}
 \end{equation}
 Explicitly, the smooth term reads
 \begin{equation}
 f_0(E)=\frac{1}{2\pi}\frac{\exp{(-K/\hbar)}}{\sqrt{(-1)^{d-1}\det(W-I)}}
 \label{cw3}
 \end{equation}
 where $K$ is the (imaginary) action of the periodic instanton (that is the full orbit
 back and forth across the potential well), $d$ the number of freedoms of the system and
 $W$ the $2(d-1)\times 2(d-1)$ stability matrix of the instanton.
 The oscillatory term is  not explicitly written by Creah and Whelan, but it is rather
 simple to calculate it from their papers. In the simple case where the classical orbit which is the
 real continuation of the instanton  inside the potential well is a periodic orbit (this is for example
 the case when 
 the instanton is along a symmetry axis of the potential), it is not surprising that the oscillatory
 terms will be governed by the properties of this real periodic orbit. Indeed, it is known
 that the eigenstates inside the well will be scarred by such an orbit, thus showing
 either an increased or decreased probability density at the point where the instanton emerges.
 It is thus reasonable to expect that scarred (resp. anti scarred) states will show an increased
 (resp. decreased) tunneling probability. The modulations of the weighted density
 of states are thus related to the action of the real periodic orbit. More specifically, one
 gets:
  \begin{equation}
 f_{\mathrm{osc}}(E)=\frac{1}{\pi}\ {\mathrm{Re}}\ \sum_{j=1}^{\infty}
 {\frac{\exp{\left[(-K+ijS)/\hbar\right]}}{\sqrt{(-1)^{d-1}\det(WM^j-I)}}}
 \label{cw4}
 \end{equation}
where $S$ is the (real) action of the periodic orbit in the well and $M$ its stability matrix.
The sum over $j$ just takes into account the repetitions of this orbit. This approach
is restricted to a low tunneling rate, when repetitions of the instanton
give negligible contributions. The fact that, in the contribution of the instanton,
there is a $1/2\pi$ prefactor, half the prefactor for the oscillatory term,
is not trivial, but explained in ref.~\cite{creagh96a}.

\section{The hydrogen atom in parallel fields}

We consider a hydrogen atom placed in static parallel magnetic and electric
fields. The Hamiltonian of the system is (for infinite mass of the nucleus,
neglecting relativistic and QED corrections, in atomic units)

\begin{equation}
H=\frac{{\bf p}^2}{2}-\frac{1}{r} - F z +
\frac{\gamma^2}{8}(x^2+y^2)+\frac{\gamma}{2}L_z
\label{ham}
\end{equation}
where $\gamma$ stands traditionally for the magnetic field in atomic units
($\approx 2.35\times
 10^5$ Tesla)
while $F$ is the static electric field (in atomic units of 
$\approx 5.1\times 10^{11}$ V/m)
assumed to be oriented, together with
the magnetic field, along the $Oz$ axis. The system obeys cylindrical symmetry
and $L_z$ is a constant of motion. The last, Zeeman term
 in the Hamiltonian gives thus a constant shift (for a given $L_z$) and
will be omitted for simplicity.

Classically, the atom may ionize for energies $E_{\rm cl}>-2\sqrt{F}$.
 Note that
ionization occurs in the $z$ direction - the diamagnetic term
provides a two-dimensional harmonic oscillator  binding potential in the
perpendicular directions.

The character of the classical motion depends on the energy as well as on
the relative magnitude of electric and magnetic fields, as discussed long
time ago in~\cite{cacciani86,waterland87}.
In fact the system obeys
the standard classical scaling laws \cite{delande89a,friedrich89}.
Explicitly,  scaling with respect to
the magnetic field as $\tilde {\mathbf r}=  {\mathbf r}\gamma^{2/3}$, 
$\tilde {\mathbf p}= {\mathbf p}\gamma^{-1/3}$,
$\epsilon=E\gamma^{-2/3}$, $f= F\gamma^{-4/3}$, $\tilde L_z= L_z\gamma^{1/3}$,
leads to a new Hamiltonian dependent on two parameters only,
the scaled energy $\epsilon$ and scaled electric field $f$
\begin{equation}
\tilde H=\frac{{\bf \tilde p}^2}{2}-\frac{1}{\tilde r} - f \tilde z +
\frac{1}{8}(\tilde x^2+\tilde y^2)=\epsilon.
\label{hams}
\end{equation}
Later on we shall drop the $\tilde{}$ sign using classically scaled variables only.

With this scaling and for $f=0$ (pure magnetic field), 
the motion is predominantly regular for
small $\epsilon<-0.5$; for larger $\epsilon $ a gradual transition to chaos
takes place so that, for $\epsilon>-0.12,$ practically all the available
phase space becomes chaotic~\cite{delande89a,friedrich89}.
 This character of the motion is
basically preserved for $f>0$, provided
\begin{equation}
\epsilon < \epsilon_{\mathrm{ion}} = -2\sqrt{f},
\label{classth}
\end{equation}
  i.e., below the
classical ionization threshold.

Quantum mechanics does not preserve the scaling. Instead of finding, however,
eigenenergies at given values of
 magnetic $\gamma$ and electric $F$ fields,
it is a celebrated tradition now to consider scaled spectra
\cite{delande89a,friedrich89},
i.e. choose values of external fields as to obtain
eigenenergies at fixed $\epsilon$. This procedure is straightforward.
Rewriting the Schr\"odinger equation, one may obtain a generalized eigenvalue
problem for fixed $\epsilon$  (and $f$ in our case) from which quantized
field values $\gamma_n^{-1/3}$ are obtained. If one were to get back to
the original problem, then a given $\gamma_n$ value together with the definition of
$\epsilon$  yields the energy $E_n$ which is an eigenvalue of the original
Schr\"odinger equation for that $\gamma_n$ field value. The set of
$\gamma_n$ obtained for a fixed values of $\epsilon$ and $f$ corresponds
to the very same classical dynamics while different $\gamma_n^{-1/3}$ play
the role of different values of the effective Planck constant  $\hbar_{\mathrm{eff}}$.

One may then expect that to study quantum tunneling in the semiclassical regime
with a well defined classical mechanics, it is sufficient to diagonalize a
standard scaled problem.  This is, however, not completely true: indeed, because
tunneling implies that the electron ionizes, the energy spectrum is not a discrete spectrum
of bound states, but rather composed of resonances. Far below the classical
ionization threshold, the widths of the resonances are extremely small and
can be neglected. Then diagonalization of the Hamiltonian in a convenient basis set
may produce a discrete energy spectrum which very well approximates
the true resonances. On the other hand, far above the ionization threshold,
the spectrum is continuous and basically unstructured.
We are interested in the intermediate situation, in the vicinity
of the classical ionization threshold, where the resonances have
a small but significant width due to tunneling followed by ionization.
The treatment of tunneling resonances necessitates
a further standard extension, known from the pure magnetic field case above
the ionization threshold \cite{gremaud93,dupret95}:
a complex rotation approach \cite{reinhardt82,maquet83,ho83}. The idea is to
apply the following complex scaling (or rotation if viewed in the complex plane):
 $\mathbf{r} \rightarrow \mathbf{r}\exp(i\Theta)$,
 $\mathbf{p} \rightarrow \mathbf{p}\exp(-i\Theta)$ to the Hamiltonian of the
 system, where $\Theta$ is a real positive parameter representing the complex rotation angle
 (typically of the order of 0.1 rad). The transformed Hamiltonian is no longer an Hermitian operator,
 and its diagonalization yields complex eigenvalues $E_n-i\Gamma_n/2,$
 where $E_n$ is the energy of the resonance and $\Gamma_n$ its width,
 i.e. the inverse of its lifetime. 
Well below the classical ionization threshold, eq.~(\ref{classth}), 
the widths should be vanishingly small; with increasing $f,$  some
states, notably those extended in $z$ direction, i.e., along the
``ionization direction'', should have 
increased imaginary parts indicating
tunneling ionization. Above the threshold, eq.~(\ref{classth}),  
ionization
becomes classically allowed and the widths are expected to be large. 
Moreover, in the tunneling
regime, the widths should on the average exponentially decrease in magnitude
with decreasing $\hbar$ according to an $\exp(-K/\hbar)$ law with $K$
being a characteristic tunneling action.

\section{Numerical results - Shift of the effective ionization threshold}

The expectations described in the previous section are based on a rough
classical analysis of the ionization process. 
In order to test these ideas and the semiclassical prediction for the
widths of the resonances, we have performed extensive numerical studies
of the energy spectrum of the system. The matrix representing the 
complex rotated Hamiltonian in a Sturmian basis set \cite{delande86}
is easily obtained, as matrix elements have strong selection rules
and are all known as simple analytic expressions. 
The matrix in then diagonalized using the Lanczos algorithm \cite{Lanczos},
producing several hundreds or thousands fully converged eigenvalues.
We have carefully checked that all eigenvalues presented in this paper
are fully converged. The only limitation is that the calculation is performed
in double precision, yielding about 12 significant digits. This also implies
that widths (tunneling rates) smaller than $10^{-12}$ cannot be accurately computed.

It turns out that, below the classical ionization threshold, eq.~(\ref{classth}), 
the widths of the resonances are usually very small. Moreover, as we are
interested in the situation when the classical dynamics inside the potential
well is chaotic, we have to use a rather large value of the scaled energy
$\epsilon$ -- typically $\epsilon=-0.1$ -- which in turns
correspond to a rather small value of the scaled electric field
at the ionization threshold, that is typically $f=0.0025$ from eq.~(\ref{classth}).
For these values, we observed that the numerically computed widths are all
vanishingly small, smaller than the $10^{-12}$ accuracy of the numerics.

This can be understood from eq.~(\ref{cw3}). Indeed, the stability exponent of
the instanton orbit is in our specific case enormous. The reason is that the potential
in the vicinity of the saddle point is very anisotropic. 
It is strongly bounded by the diamagnetic term in the
transverse $(x,y)$ plane but has a only a very smooth potential maximum in the
field $(z)$ direction. The instanton can be seen as a real orbit propagating in the
inverted potential. This inverted potential has a shallow minimum in the $z$ direction
but falls down very rapidly in the transverse directions: the instanton moves along
a sharp potential ridge and is thus extremely unstable. We show in the appendix
how to calculate the action $K$ and stability matrix $W$ of the instanton. For small
$f$ (the regime we are interested in), the following approximate expressions 
are sufficient:
\begin{equation}
K(\epsilon) = - \frac{2\pi(\epsilon-2f^{1/2})}{2^{1/2}f^{3/4}}
\label{K}
\end{equation}
and
\begin{equation}
(-1)^{d-1} \det(W-I) = \exp{\left(\frac{\pi}{2^{1/2}f^{3/4}}\right)}
\label{denum}
\end{equation}

For $f=0.0025,$ the denominator in eq.~(\ref{cw3}) is thus of the order of $10^{-43},$
which explains that the widths cannot be measured in a numerical experiment
\footnote{This effect was not observed in the various numerical experiments of
Creagh and Whelan. This is because, in their case, the potential varies
quite rapidly along the instanton trajectory, having a shallow minimum in the transverse direction.
This results in the denominator being of order unity, several tens of orders of magnitude
larger than in our case.}.

An alternative, equivalent, ``quantum" explanation can be given. 
The magnetic field term in the Hamiltonian is responsible for a  harmonic 
potential in the direction perpendicular to the fields. Thus the
quantum mechanical energy of the motion in the $x-y$ plane cannot
be smaller than the energy of the lowest state of the corresponding
oscillator. For $L_z=m=0$ the energy in question is the ground state
energy $E_0=\gamma/2$ while for other (conserved) $m$ values it is
$E_m=(|m|+1)\gamma/2$. Reaching the energy of
the classical ionization threshold is thus not sufficient for the quantum system
to ionize. It requires the additional zero-point energy $E_m$ to be able to
overcome the potential barrier. As the harmonic potential is quite strong,
this excess energy is rather high and has the effect of tremendously reducing the
ionization probability. For the scaled problem, the energy shift $E_m$ translate
into a shift of the scaled energy:
\begin{equation}
\epsilon_m = E_m \gamma^{-2/3} = \frac{|m|+1}{2} \gamma^{1/3} = \frac{|m|+1}{2} \hbar_{\mathrm{eff}}.
\label{shift}
\end{equation}

The equivalence of the two points of view can be established by noting that the
expression~(\ref{denum}) has itself an exponential dependence, which can be combined
with the numerator in eq.~(\ref{cw3}). We obtain:
\begin{equation}
f_0(E)= \frac{1}{2\pi} \exp{\left[\frac{2\pi(\epsilon-2f^{1/2}-\hbar_{\mathrm{eff}}/2)}{\hbar_{\mathrm{eff}} 2^{1/2}f^{3/4}}\right]}
\end{equation}
which can also be written as:
\begin{equation}
f_0(E) = \frac{1}{2\pi} \exp{\left[\frac{-K(\hat{\epsilon})}{\hbar_{\mathrm{eff}}}\right]}
\label{f0}
\end{equation}
where (here for $m=0$)
\begin{equation}
\hat{\epsilon} = \epsilon - \epsilon_m.
\label{hat_epsilon}
\end{equation}
The physical meaning of these equations is rather clear. In effect, tunneling
can be described with a standard exponential[action/$\hbar_{\mathrm{eff}}$],
 provided the amount $\epsilon_0=\hbar_{\mathrm{eff}}/2$ of energy 
 is subtracted from the total available energy
$\epsilon.$ The global effect of the degrees of freedom transverse to the
instanton is nothing but a shift of the  energy available for tunneling by the zero-point energy
in the transverse direction.

Such an analysis does not take into account the azimuthal symmetry
around the fields axis and the fact that the contributions of the various $m$ values
can be separated in the numerical calculation. Not surprisingly, tunneling
is much more effective for $m=0$ states which are not repelled from the $z$ axis by a centrifugal potential
and thus feel more efficiently the instanton. In the quantum point of view, such states
have the lowest transverse zero-point energy $\epsilon_0.$ For other $m$ values, the treatment
of Creagh and Whelan has to be adapted: instead of considering the full semiclassical Green function
of the system, one must expand it on the various $m$ subspaces and use only the relevant component
in each subspace. A similar problem occurs in periodic orbit theory when one is
interested in the contribution to the density of states of an orbit along the $z$ axis.
How to deal with such a problem has been described in a general manner by Bogomolny~\cite{Bogomolny88b,Bogomolny89}
and in a specific case by Shaw and coworkers~\cite{shaw95}. The rule is that higher powers
of the stability matrix come into play, for example $M^{|m|+1}$ instead of $M$ for the real orbit. 
The situation is similar for the instanton, resulting in the denominator
being raised to power $|m|+1.$ The net effect is again taken into account by 
shifting the scaled energy by $\epsilon_m=(|m|+1)\epsilon_0,$ i.e. the transverse zero-point
energy in the $m$ subspace. Thus results in an effective quantum ionization threshold:
\begin{equation}
\epsilon_{\mathrm{ion}}^{q} = -2\sqrt{f} + \frac{|m|+1}{2} \hbar_{\mathrm{eff}}
\label{qth}
\end{equation}

\begin{figure}
\psfrag{=1/hbareff}{\Huge{$=1/\hbar_{\mathrm{eff}}$}}
\centering
{\includegraphics[angle=-90,width=8cm]{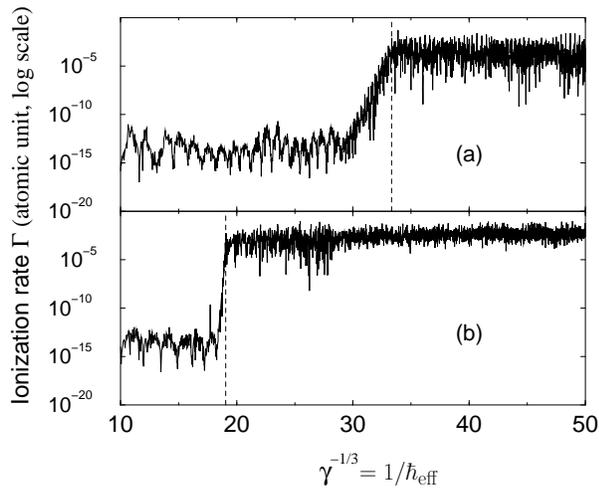}}
\caption{
 The ionization rates (widths) of resonances
 of the hydrogen atom in parallel electric and magnetic fields,
  as a function of the effective principal quantum number 
$\gamma^{-1/3}=1/\hbar_{\mathrm{eff}}$ which plays the role
of the inverse of the effective Planck's constant. The  diagonalization is performed
at fixed value of the scaled energy $\epsilon$ and yields
complex eigenvalues, their real part corresponds to quantized values of
magnetic field $\gamma$ for fixed values of $\epsilon$ and $f$. (a)
shows data obtained for angular momentum $m=0$, scaled electric field $f=0.0025$
and $\epsilon=-0.085$, i.e. above the classical ionization threshold as
given by condition~(\protect{\ref{classth}}). Observe very small ionization
widths for $\gamma^{-1/3}<28$, then a rapid, on average increase and saturation
for  $\gamma^{-1/3}>33.3$. The region of rapid increase corresponds
to tunneling as explained in the text. The dashed vertical line gives
the quantum ionization threshold, eq.~(\protect{\ref{qth}}). (b) shows
the data for $m=3$, $f=0.0025$, and $\epsilon=0.005$ showing a similar
behavior. Now the threshold is at 
$\gamma^{-1/3}\approx 19$.
}
\label{scaled}
\end{figure}

Because of the transverse zero-point energy, in the presence of magnetic field, {\it larger}
electric field strengths are necessary to observe the same
ionization yield, or, conversely, larger scaled energy is required for a fixed electric field
strength. We have thus performed numerical diagonalization of the scaled
Hamiltonian {\it above} the classical ionization threshold, eq.~(\ref{classth}). 
The results is shown in Fig.~\ref{scaled}, where the widths of the resonances
are plotted versus the quantized value of $\gamma^{-1/3}=1/\hbar_{\mathrm{eff}},$ the
inverse of the effective Planck's constant. At low  $\gamma^{-1/3},$ i.e. large $ \hbar_{\mathrm{eff}},$
the transverse zero-point energy is so large that the quantum ionization threshold, eq.~(\ref{qth}),
is far above the scaled energy of the state which consequently has a vanishingly small
ionization rate. This corresponds to the region
$\gamma^{-1/3}<28$ in Fig.~\ref{scaled}(a), where the widths are smaller than the numerical
accuracy. As  $\gamma^{-1/3}$ is increased, the quantum ionization threshold decreases
and significant tunneling takes place, as observed in the range $28<\gamma^{-1/3}<35.$
Finally, at sufficiently high $\gamma^{-1/3}$ value, the scaled energy is higher than
the quantum ionization threshold and direct ionization takes place. There, the ionization
rates are large, comparable to the spacing between consecutive resonances and the tunneling
regime is left. In the figure, the $\gamma^{-1/3}$ value where the
quantum ionization threshold is reached is marked by the dotted line, 
and agrees with the numerical results.
The two values of
$\epsilon$ for $m=0$ and $m=3$ have to be chosen quite different
in order to observe
the transition within the range of $\gamma^{-1/3}$ available
from numerical diagonalization. Let us note that also in the pure
magnetic field case, the statistics of level spacings in the vicinity
of the
ionization threshold is sensitive to the very same quantum
threshold law \cite{kuba95a,dupret95}.

\section{Numerical results at constant modified scaled energy}

The behavior observed in Fig.~\ref{scaled} has important consequences.
To study tunneling, we should consider only the region just below
the threshold; this region is very small and
the tunneling rate changes very rapidly with $\gamma$. Thus, 
scaled spectroscopy is not appropriate  for the analysis of statistical
properties of tunneling. As it is clear from the discussion above
and the examples depicted in the figures, the proper parameter
characterizing the spectrum is not $\epsilon$ but rather
$\hat{\epsilon}=\epsilon-\epsilon_m$. 
In order to overcome the difficulty described in the previous section,
a simple solution is thus to scale the problem following the effective quantum
ionization threshold instead of the classical one. One then gets rid of the
huge denominator due to the transverse motion and may more easily
concentrate on the interesting dynamics, namely the interplay between
the instanton and the chaotic dynamics inside the potential well.
We will thus solve the Schroedinger equation, not at constant scaled
energy $\epsilon$, but at constant modified scaled energy $\hat{\epsilon}$,
eq.~(\ref{hat_epsilon}). This results in the following 
generalized (non linear) eigenvalue problem for the effective
Planck's constant $\hbar_n$ and
the eigenstates $\phi_n(\mathbf{r}):$
\begin{equation}
\left(-\frac{\hbar_n^2}{2} \Delta + \hbar_n \frac{|m|+1}{2}  - \hat{\epsilon}
-\frac{1}{r} - f z +
\frac{x^2+y^2}{8}\right) \phi_n(\mathbf{r}) = 0
\end{equation}
with $\Delta$ the Laplace operator.

\begin{figure}
\psfrag{=1/hbareff}{\Huge{$=1/\hbar_{\mathrm{eff}}$}}
\centering
{\includegraphics[angle=-90, width=8cm]{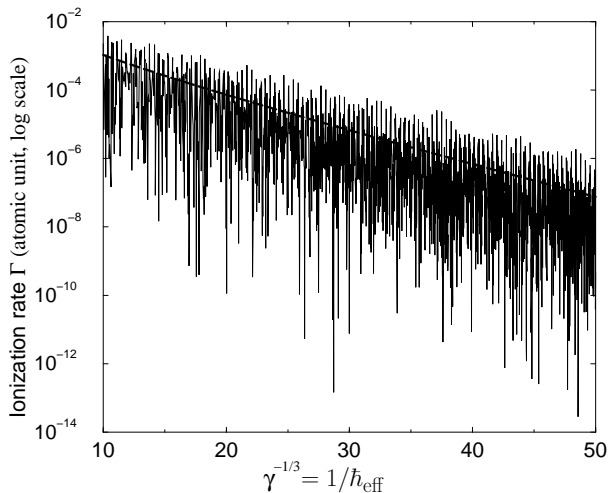}}
\caption{
Widths (ionization rates) of the resonances of the hydrogen atom in parallel
electric and magnetic fields, computed at constant
modified scaled energy, eq.~(\ref{hat_epsilon}), $\hat{\epsilon}=-0.1005$, 
i.e. $0.0005$ below the quantum 
ionization threshold, eq.~(\protect{\ref{qth}}). Parameters are $m=0$, $f=0.0025.$
The widths (in logarithmic scale) are plotted versus 
$\gamma^{-1/3}=1/\hbar_{\mathrm{eff}}.$  
The data show the exponential decrease of the rate for
$\hbar_{\mathrm{eff}}\rightarrow 0$ characteristic for tunneling process.
The dashed line is the average behavior predicted from 
the semiclassical analysis, eq.~(\protect\ref{f0}).
One may also notice periodic short-range modulations of the ionization rates
(with a period close to 0.4); this is a manifestation of scarring
by the periodic orbit along the fields axis and is discussed in section~\protect\ref{modulations}.
}
\label{unscaled}
\end{figure}

This equation is solved by expansion over a Sturmian basis and a modified version
of the Lanczos algorithm adapted to such a generalized eigenvalue problem~\cite{numrec}. 
We have been able to
obtain a few thousands of resonance widths for a given $m$ value, all
lying in the tunneling regime. An example is presented in Fig.~\ref{unscaled}.  

As expected from eq.~(\ref{f0}), the ionization rate shows an overall exponential
decrease with $1/\hbar_{\mathrm{eff}}.$ The rate of this decrease is directly
related to the tunneling action of the instanton: the prediction of eq.~(\ref{f0})
is shown as a dashed line in the figure. Obviously, the agreement is excellent.
It should be noted that the semiclassical prediction is entirely obtained
from classical ingredients and free of any parameter. Note also the existence
of very large fluctuations -- characteristic of chaotic tunneling -- around the mean value. 

In order to make a more quantitative test, we remove the global exponential decrease
and define, following~\cite{creagh99a}, a rescaled width:

\begin{equation}
y_n = \frac{\rho_0(E_n)}{f_0(E_n)}\Gamma_n.
\label{y}
\end{equation}
where $\rho_0(E)$ is the density of states.

\begin{figure}
\psfrag{=1/hbareff}{\Huge{$=1/\hbar_{\mathrm{eff}}$}}
\centering
{\includegraphics[angle=-90, width=8.0cm]{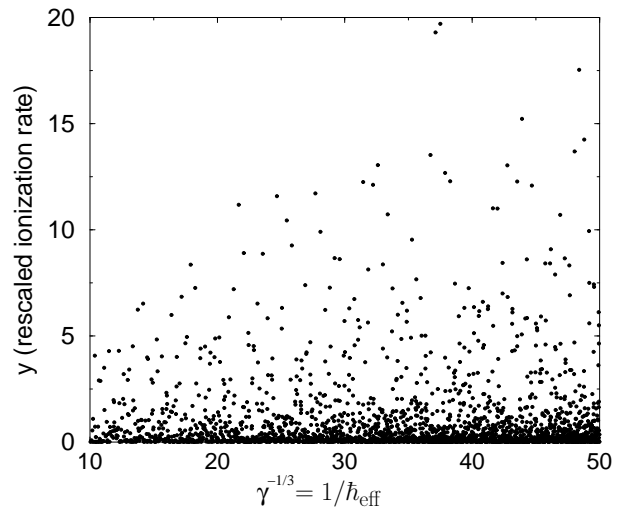}}
\caption{
Rescaled widths (ionization rates) $y_n,$ eq.~(\protect\ref{y}), 
of the resonances of the hydrogen atom in parallel
electric and magnetic fields (same data as in Fig.~\protect\ref{unscaled}).
As expected from the semiclassical analysis, the $y_n$ values
have an average value close to unity, with large fluctuations.
Note especially the large proportions of very small widths, 
characteristic of chaos assisted tunneling. 
}
\label{rescaled}
\end{figure}

From its definition, the $y_n$ should have average value unity in the semiclassical
limit. The distribution of the $y_n$ (same data than in Fig.~\ref{unscaled}) is shown
in Fig.~\ref{rescaled}, on a linear scale. It has very large fluctuations -- several orders
of magnitude with a large proportion of very small ionization rates -- but we checked
that the average value of $y_n$ is constant across the spectrum within a few percent
(although the $\Gamma_n$ themselves vary over five orders of magnitude) and equal to
$0.95 \pm 0.03.$ This is only slightly smaller than unity. The difference
might be due to deviations from harmonicity of the potential in the vicinity
of the saddle point (an assumption made in our calculation). Another plausible
source of deviation is the assumption made in the calculation of Creagh and Whelan
that every electron which tunnels through the barrier will eventually ionize; although this is
very likely to happen, the channel along the $z$ axis may also reflect a small part
of the electronic wavefunction, even after tunneling took place. This would manifest itself
by the $y_n$ being smaller than unity.

The main point remains that semiclassics is able to predict quantitatively the
average behavior of the ionization rates in the tunneling regime.

\section{Fluctuations of the ionization rates -- Effect of scarring}
\label{modulations}
\begin{figure}
\centering
{\includegraphics[angle=-90,width=8cm]{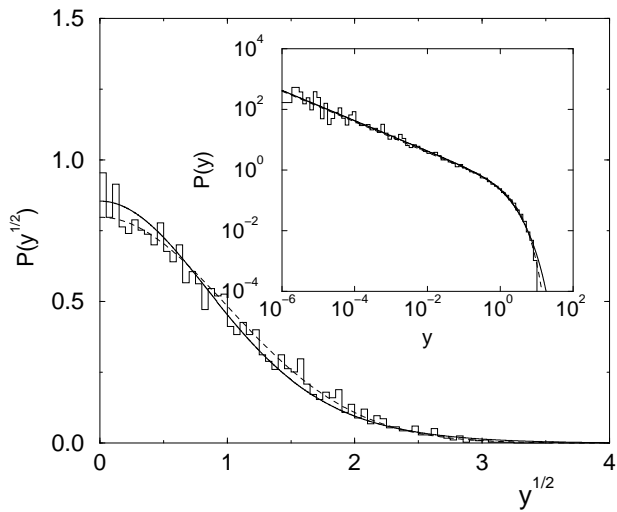}}
\caption{
Statistical distribution of the rescaled ionization rates (widths) $y$
of the resonances of the hydrogen atom in parallel
electric and magnetic fields. The data are taken
at fixed modified scaled energy $\hat{\epsilon},$ eq.~(\protect\ref{hat_epsilon}),
and rescaled according to eq.~(\protect\ref{y}) in order to take into account the
exponential decrease due to the tunneling through the potential barrier.
These data are obtained for the $m=3$ series at scaled electric field $f=0.0025$.
In order to improve the statistics, several distributions obtained
for various values of $\hat{\epsilon}$ slightly below -0.1 are used.
The inset shows the distribution $P(y)$ on a double logarithmic scale,
with a $y^{-1/2}$ behavior at small $y$ characteristic of ionization
with only one open channel, and a exponential tail at large $y.$
The numerical results are shown by the
histogram, the dashed line is the Porter-Thomas distribution,
eq.~(\protect\ref{pt}) predicted by Randon Matrix Theory, while
the solid line is the prediction of eq.~(\protect\ref{modified_pt})
obtained by taking into account the scarring of the eigenstates
by the orbit along the field axis. Note that there is no adjustable
parameter. The main figure shows the distribution of $\sqrt{y}$,
which is not singular as $y\to 0,$
on a double linear scale. The prediction of Random Matrix Theory is
a pure Gaussian.
For the $m=3$ series, the effect of scarring is
small and both theoretical distributions agree well with the 
numerical results.
}
\label{stat-m3}
\end{figure}

Beyond the average behavior discussed in the previous section, we are also interested in the
fluctuations of the ionization rates. The most probable origin of these fluctuations
is the fact that the classical dynamics inside the potential well is chaotic. 
This implies that the resonance wavefunctions in the well display apparently erratic fluctuations
from state to state. States with a large probability density
near the classical saddle point are more likely to tunnel and ionize that the ones
with small probability density. As a simple approximation, the tunneling probability
and thus the ionization rate is proportional to the squared overlap between the
eigenstate and a wavepacket optimally tuned for tunneling, i.e. built to follow
the instanton trajectory. Creagh and Whelan have shown how to explicitly build such a
wavepacket~\cite{creagh02}. For a quantum chaotic system, the simplest model
for describing the statistical properties of the energy spectrum and eigenstates is to use
Random Matrix Theory (RMT)~\cite{brody81,mehta91}. There, it is assumed that any unknown matrix element will be statistically
described by a Gaussian distribution. In our case, although the system is not time-reversal invariant
(because of the magnetic field), it has a generalized time-reversal symmetry and the
Gaussian Orthogonal Ensemble (GOE) of random matrices must be used. Thus the matrix element
is purely real and its square, and consequently the ionization rate, will be described
by a Porter-Thomas distribution~\cite{porter56,brody81}:
\begin{equation}
P(y)=\frac{1}{\sqrt{2\pi y\bar{y}}}\exp(-y/2\bar{y})
\label{pt}
\end{equation}
where $\bar{y}$ is the mean value of $y$, unity in our case.
In Fig.~\ref{stat-m3}, we show the numerically obtained distribution for the $m=3$ series
compared with the Porter-Thomas prediction, on a double-logarithmic scale which is more convenient
to display the large fluctuations. The agreement is excellent, which proves that the distribution
of ionization rates in our realistic problem can be quantitatively predicted, using a combination
of semiclassics (for the mean value) and Random Matrix Theory (for the fluctuations).

\begin{figure}
\centering
{\includegraphics[angle=-90,width=8cm]{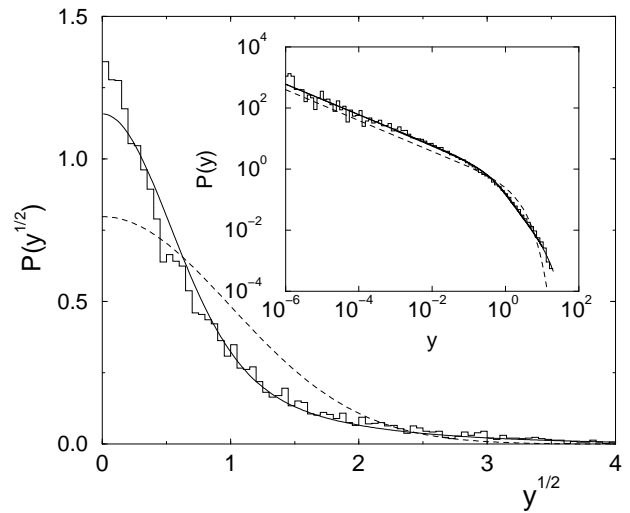}}
\caption{
Same as figure~\protect\ref{stat-m3}, but for the $m=0$ series.
The effect of scarring is
much more important and large deviations from the
Porter-Thomas (Random Matrix Theory) distribution (dashed line)
are observed. In contrast, the agreement with the
model taken into account the scarring by the $z$ orbit, 
eq.~(\protect\ref{modified_pt}) and solid line,
is much better. This proves that our model -- with no adjustable
parameter -- describes properly the physics of tunneling
and ionization of chaotic states in the presence of scarring.
}
\label{stat-m0}
\end{figure}

\begin{figure}
\centering{{\includegraphics[angle=-90, width=8cm]{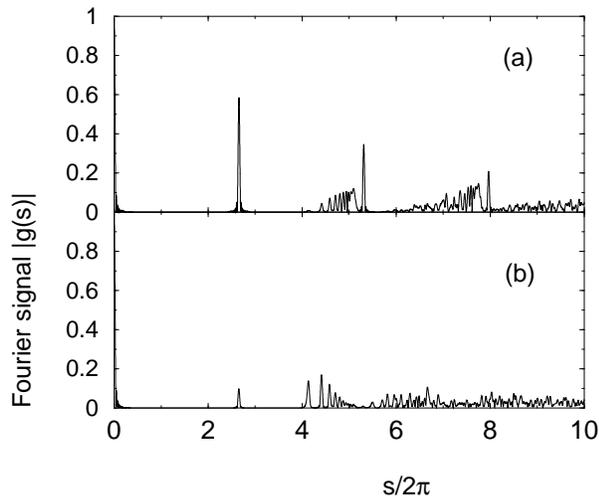}}}
\caption{
Fourier transforms of the distribution of widths (ionization rates)
with respect to $1/\hbar_{\mathrm{eff}},$ eq.~(\protect\ref{gs}),
obtained for $f=0.0025$ and $\hat{\epsilon}=-0.1005$ in the $m=0$ (a)
and $m=3$ (b) series.
It displays a large peak (hardly visible) at $s=0$ with amplitude unity,
as expected from semiclassics; this peak describes the average
behavior of the widths. The other peaks correspond to modulations
of the width associated with periodic orbits inside the inner potential well.
The peaks are much more pronounced for the $m=0$ series (a) than for the
$m=3$ series (b), because the former are more strongly
scarred by the orbit along the fields axis, with action $S_{\mathrm{clas}}/2\pi=2.655.$
The repetitions of this orbit are clearly visible, with amplitudes
forming a geometric series. The heights of these peaks are directly related
to the instability of the periodic orbit along the fields axis. In (a),
the amplitude of the first peak is $0.589\pm0.005$ in excellent
agreement with the semiclassical prediction $0.592,$ which involves
exclusively classical properties of the periodic orbits.
The other peaks, appearing at actions clustered slightly below
the repetitions of the main orbit, correspond to other
orbits in the inner potential well, which are slightly off the $z$ axis
but approach the saddle point. In (b), the 
centrifugal potential prevents the periodic orbits from
strongly scarring the quantum states, and the amplitudes of the peaks 
(especially the ones associated with repetitions of the $z$ orbit) are
much smaller.
}
\label{ft}
\end{figure}

The results for the $m=0$ series are shown in Fig.~\ref{stat-m0}. The overall agreement is 
rather good, with a clear $y^{-1/2}$ behavior at small $y$ and an exponentially
small tail at large $y.$ However, a significant deviation is clearly visible at intermediate
values. What is the origin of this deviation ? We have been able to show that it is directly related
to the unstable periodic orbits inside the chaotic potential well. The quantitative interpretation is based
on the semiclassical prediction, eqs.~(\ref{cw1}-\ref{cw4}). A clue is provided by a careful inspection
of Fig.~\ref{unscaled} which shows that fluctuations around the average trend are not random but clearly
display a short-range oscillation (about 100 oscillations in the covered range). 
The simplest way of measuring this oscillations is to perform a Fourier transform with respect
to $1/\hbar_{\mathrm{eff}},$ a standard tool in periodic orbit theory. We define:
\begin{equation}
g(s) = \frac{2 \pi}{\Delta} \sum_n{\Gamma_n \exp{(K/\hbar_n)}  \exp{(-i s/\hbar_n)}}
\label{gs}
\end{equation}
where the sum is taken over some range of $1/\hbar$ of length $\Delta.$
The function $g(s)$ is shown in Fig.~\ref{ft} both for $m=0$ and $m=3.$
As expected, $g(0)$ is very close to unity (this proves that the actual average
width is well predicted by the semiclassical formula, eq.~(\ref{f0}).
$g(s)$ has  a very large peak around $s/2\pi=2.655,$  with harmonics at integer
multiples, but also smaller peaks at other values.
From eq.~(\ref{cw4}), the peaks are expected to take place at the actions of the
periodic orbits inside the well which are real continuations of the imaginary instanton.
In our case, this orbit is entirely along the $z$ axis and its classical action 
$S_{\mathrm{clas}}$ is easily 
computed. We find $S_{\mathrm{clas}}/2\pi=2.655$ for the parameters of the figure,
in perfect agreement with the numerical quantum
calculation. It should be noticed that we use for the classical calculation the scaled
energy -0.1005, i.e. the value of the {\it modified} scaled energy of the quantum calculation
\footnote{For the $\hbar_{\mathrm{eff}}$ values used in our calculation, the
scaled energy is far {\it above} the classical threshold.}. The fact
that both agree validates our correction and fully confirms the important role
of the zero-point transverse energy.

\begin{figure}
\centering{{\includegraphics[angle=-90, width=8cm]{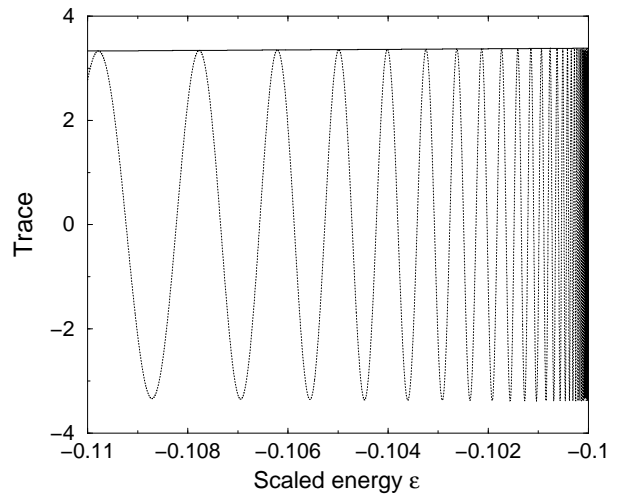}}}
\caption{
Trace of the stability matrix $M$ of the periodic orbit
along the fields axis, as a function of the scaled energy (dotted line).
The orbit is stable when the trace is in the [-2,2] range, unstable otherwise.
There is a series of stable-unstable bifurcations accumulating at the
the saddle point energy $\epsilon=-0.1$ ($f=0.0025$ in this plot).
In contrast, the contribution $|\sqrt{-\det(WM-1)}/\sqrt{-\det(W-1)}|$, plotted
as a solid line, is a smooth function of $\epsilon.$ It is precisely
this contribution which enters the denominator of the semiclassical expansion,
eq.~(\protect\ref{cw4}).  
}
\label{trace}
\end{figure}

The semiclassical formula~(\ref{cw4}) also predicts the amplitude of the peak that
should be observed in the Fourier transform $g(s).$ There is however an important subtlety
here. The monodromy matrix of the real orbit along the $z$ axis enters the formula.
It turns out that, because the orbit is very close to the saddle point (reached at
$\epsilon=-2\sqrt{f},$ it undergoes an infinite series of bifurcation as
$\epsilon \to -2\sqrt{f}.$ At closely spaced $\epsilon$ values, the orbit
loses and regains stability. At each bifurcation, a new periodic orbit is born,
which is off the $z$ axis, but close to it. Such a phenomenon is well known when a particle
either approaches a saddle point (see for example the Henon-Heiles
model in~\cite{brack01}) or explores a channel with a long-range potential,
as for example is the case for a Rydberg series converging to an ionization threshold~\cite{wintgen87}.
We show in Fig.~\ref{trace} the trace of the stability matrix of the $z$ orbit as a function of the
scaled energy $\epsilon$ for $f=0.0025,$ which clearly shows this series of stable-unstable bifurcations.
However, we also plot on the same figure the denominator of the semiclassical contribution, eq.~(\ref{cw4}),
to the ionization rate, whose calculation is detailed in the Appendix.
The fact that what counts is not the real orbit itself but its combination
with the instanton completely eliminates the series of bifurcations and gives a smooth contribution
as the scaled energy is varied, as observed in the numerical quantum experiment. 
Moreover, the semiclassical formula~(\ref{cw4}) predicts for $g(s)$
a peak at $S_{\mathrm{clas}}/2\pi=2.655$ 
with amplitude 0.592, while the numerical result is $0.589\pm0.005,$ in excellent agreement.
Similarly, the harmonics of the peak form approximately a geometric series with amplitude
$\approx 0.59^j$ for the $j^{\mathrm{th}}$ repetition of the primitive real orbit. The physical
interpretation is clear: because the motion in the inner potential well is chaotic, each time
the quantum particle leaves the vicinity of the saddle point (along the $z$ axis), it explores
some part of the chaotic phase space and roughly only 59\% of the electronic density
is reflected by the nucleus back along the $z$ axis.

It is important to remark that the oscillations of the widths, 
eq.~(\ref{cw4}), induced by the
orbit along the $z$ axis and all its repetitions add {\it coherently}. 
Indeed, if we assume
for simplicity that the $j^{\mathrm{th}}$ repetition contributes 
with amplitude $R^j$
(with $R\approx 0.59$ in our case) and phase $j\phi$, the series, 
including the smooth term $f_0$ can be summed exactly, leading to the
following contribution to the ionization width:
\begin{equation}
 f(\hbar)=\frac{1}{2\pi}\exp{(-K/\hbar)}
 \ \frac{1-R^2}{1+R^2-2R\cos{(S_{\mathrm{clas}}/\hbar-\phi)}}
\end{equation}
This, in turn, predicts that the average normalized widths 
are not uniformly distributed, but should follow
the distribution:
\begin{equation}
\bar{y}(\hbar) = \frac{1-R^2}{1+R^2-2R\cos{(S_{\mathrm{clas}}/\hbar-\phi)}}
\label{ysc}
\end{equation}
The physical interpretation of this distribution is simple. 
It is nothing, but the function giving the
intensity transmitted through a Fabry-Perot optical 
cavity with reflexion coefficients $R$ for the combination of the two
mirrors, phase shift $\phi$ at the reflections and optical 
length $S_{\mathrm{clas}}/\hbar$ in units of the
wavelength. This is of course a {\it periodic} function 
of the variable $1/\hbar$ with period 
$2\pi/S_{\mathrm{clas}}.$ It has maxima at 
$S_{\mathrm{clas}}/\hbar-\phi$ equal to an integer
multiple of $2\pi$ -- where the value of the function is 
$(1+R)/(1-R)$ -- and minima at half-integer multiples
where the function is $(1-R)/(1+R).$ If $R$ is large, 
the maxima are sharp peaks. The analogy with the Fabry-Perot
cavity is more than formal: it actually describes how the 
electronic density can be resonantly trapped
inside the inner potential well along the $z$ axis, resulting 
in enhanced tunneling amplitude
and ionization rate. Because the dynamics is chaotic, such a 
resonant enhancement is only partial
($R$ must be smaller than unity) resulting in scarring of the 
wavefunction rather than
complete localization.

\begin{figure}
\centering{{\includegraphics[angle=0,width=8cm]{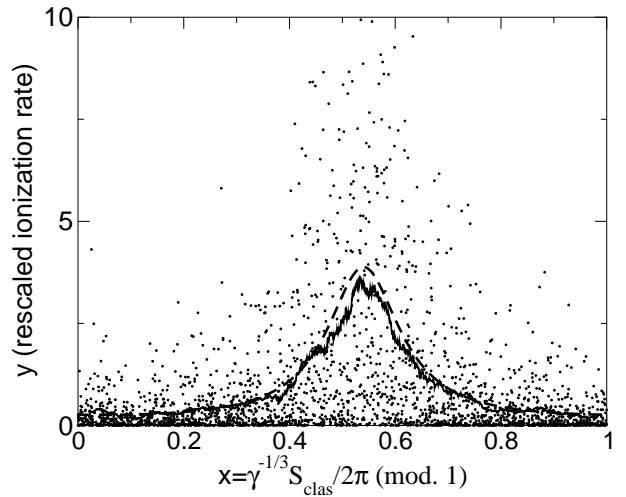}}}
\caption{\label{fp}
Rescaled ionization rates $y_n$, eq.~(\protect\ref{y}), of the
$m=0$ series for $f=0.0025$ and $\epsilon=-0.1005$, as a function
of $x,$ eq.~(\protect\ref{x}). $x$ represents
(within a multiplicative $2\pi$ factor), the phase
accumulated by the wavefunction  along the periodic
orbit in the inner potential well. There are large fluctuations,
as expected in a chaotic system, but the average behavior
is obviously dependent on $x,$ which proves
that the periodic orbit strongly affects the
width. A pure Random Matrix approach predicts
a uniform distribution independent of $x$. The solid
line is a running average which smoothes the fluctuations
and clearly shows the resonant behavior of the {\it average}
width with $x$. The dashed line is the semiclassical prediction, eq.~(\protect\ref{ybar}),
which incorporates the effect of the periodic orbit.
}
\end{figure}

In order to test whether this distribution adequately describes the numerical result, we have
``folded" all the numerical values $y_n$ inside a single ``free spectral range" of the Fabry-Perot cavity,
by plotting them against:
\begin{equation}
x_n =  \frac{S_{\mathrm{clas}} \gamma_n^{-1/3}}{2\pi}\ \ (\mathrm{mod.}\ 1) =  
\frac{S_{\mathrm{clas}}}{2\pi\hbar_n}\ \ (\mathrm{mod.}\ 1) 
\label{x}
\end{equation}
where $\alpha\ (\mathrm{mod.}\ 1)$ denotes the fractional part
of the number $\alpha.$ 
The result is shown in Fig.~\ref{fp}. Clearly the largest $y_n$ values are grouped around a well
defined $x$ value, as expected. Large fluctuations still exist; in order to smooth them, we plot
also the running average (over 100 values) which clearly presents a resonant behavior around
$x=0.54.$ The semiclassical prediction, using the $R$ value deduced from the classical
stability of the orbit, is shown as a dashed line and agrees fairly well with the numerical
result. This proves that the orbit along the $z$ axis plays the dominant role
in our problem. To be completely honest, we must mention
 that the phase $\phi=0.54\times 2\pi$, 
which is directly
related to the position of the maximum in the plot,
 has not been extracted from the classical
dynamics but fitted to the numerical data. 

The last step is to characterize precisely the fluctuations of the {\it individual} widths
that appear on top of the global exponential decrease and the modulations discussed above.
Such fluctuations are thought to be unavoidable in a chaotic system, and are even a signature of the
presence of chaos. In a semiclassical point of view, they can be seen as the 
effect of the whole set of (unstable) periodic orbits in the inner potential well.
Each orbit approaching the saddle point contributes an oscillatory term analogous to
eq.~(\ref{cw4}) to the width and the individual widths are just the result of the superposition
of plenty of such terms which oscillate rapidly: it results in apparently
random fluctuations around the mean value. A number of peaks are visible
in the Fourier transformed spectra in Fig.~\ref{ft}; semiclassical theory tells us that
they appear at the actions of periodic orbits. In our specific case, 
it is visible that most
of them are clustered at actions slightly smaller than the action 
of the $z$ orbit and its repetitions.
Physically, they correspond to orbits mainly located close to the $z$ axis, 
born from the $z$ orbit at
the bifurcations discussed above. There is actually a very large number of 
such orbits with similar
shapes, but differing by small details. Thus, it is in general difficult 
to associate a peak in the Fourier
transform with a single periodic orbit. Except for the lowest members, 
we could not assign unambiguously
such peaks. This is not a simple problem: indeed, many orbits are very 
close in phase space and,
for a finite value of $\hbar_{\mathrm{eff}},$ cannot be considered as 
isolated in the sense that the
saddle point approximation around each orbit -- a key ingredient of 
periodic orbit theory -- 
is not valid. In such circumstances, it is not possible to separate 
the contributions of the various orbits
which have to be grouped together using for example a uniform 
approximation~\cite{mouchet99}. 
Note that this is a fundamental
difficulty of periodic orbit theory, not a practical problem 
related to the proliferation of the number of orbits.

\begin{figure}
 \centering{{\includegraphics[angle=-90,width=8cm]{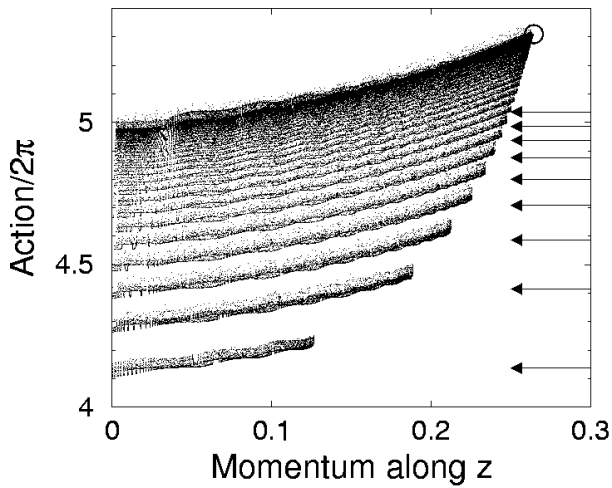}}}
\caption{\label{traj}
Distribution of actions of trajectories launched from a plane $z=z_0=8.82$
towards the saddle point as a function of the momentum along the $z$ axis
for 
$m=0$, $f=0.0025$ and $\epsilon=-0.1005$. The action is calculated along a given
trajectory till it hits the same plane also with positive momentum. Observe
a clear structure of strips. Arrows indicate actions corresponding to peaks in
the Fourier transform of the widths. The circle indicates the second repetition
of the orbit
along the $z$ axis with maximum action $2S_{\mathrm{clas}}/2\pi=5.31$ and
momentum $p_z=0.2643.$
}
\end{figure}

An interesting illustration of this problem may be obtained by launching a bunch
of trajectories from the section $z=z_0$ close to the saddle point. Following the
real trajectories  (all started with positive momentum in $z$) until they
hit again the same plane $z=z_0$ with positive momentum one can get a feeling of
the relevant dynamics. For a fully chaotic system one could naively expect that
a plot of say actions calculated along the trajectory versus the initial
momentum along the $z$ axis will not show
any structure. 
This is not true in our system as visualized in Fig.~\ref{traj}. 
Observe a strongly not ergodic behaviour with allowed actions
forming almost parallel strips. A clear accumulation of actions in strips
correlate nicely with peaks in the Fourier transform, Fig.~\ref{ft}(a), in the
range just below the second repetition at $2S_{\mathrm{clas}}/2\pi=5.31$
of the straight line periodic orbit. To each strip, apart from other
non-periodic trajectories,  several periodic
orbits contribute which slightly differ in shape (and action). As mentioned
already this makes a clear association between peaks in the Fourier transform
and given periodic orbits impossible.

The basic assumption, usual in studies of quantum chaos, is that the effect of long periodic orbits 
is to create fluctuations well described by Random Matrix Theory.
As -- see above -- the ionization rate appears as the square of some real matrix elements, the simplest
hypothesis is to assume that the fluctuations are described by a Porter-Thomas distribution, eq.~(\ref{pt}).
However, the mean value $\bar{y}$ is now taken as predicted by the semiclassical theory, i.e. eq.~(\ref{ysc}).
As the $x_n$ values are uniformly distributed, this results in a global statistical distribution:
\begin{equation}
P(y;R) = \int_0^1{\frac{1}{\sqrt{2\pi y \bar{y}(x)}} \exp{\left(-\frac{y}{2\bar{y}(x)}\right)}\ dx}
\label{modified_pt}
\end{equation}
where
\begin{equation}
\bar{y}(x) = \frac{1-R^2}{1+R^2-2R\cos{2\pi x}}
\label{ybar}
\end{equation}

This distribution is plotted in Fig.~\ref{stat-m0} as a solid line. It clearly very significantly
improves over the Porter-Thomas distribution and is in excellent agreement with our numerical data. 
Especially, it correctly describes the excess of large ionization widths.

\begin{figure}
\centering
{\includegraphics[angle=-90,width=8cm]{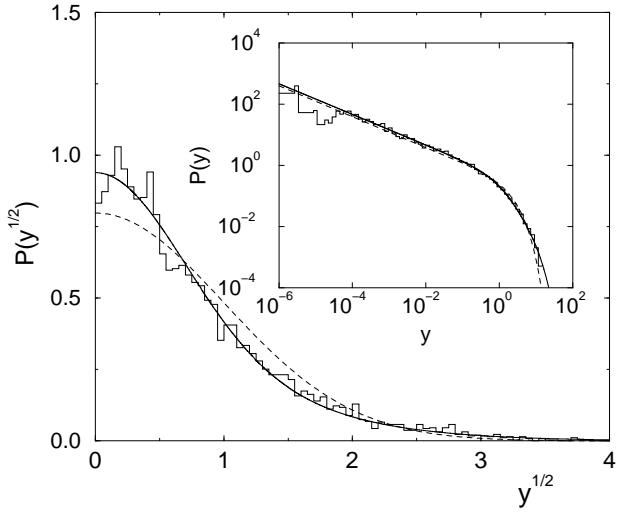}}
\caption{
Same as figures~\protect\ref{stat-m3} and \protect\ref{stat-m0}, 
but for the $m=1$ series.
The effect of scarring is
intermediate and some deviations from the
Porter-Thomas (Random Matrix Theory) distribution (dashed line)
are observed. In contrast, the agreement with the
model taken into account the scarring by the $z$ orbit, 
eq.~(\protect\ref{modified_pt}) and solid line,
is much better. 
}
\label{stat-m1}
\end{figure}

\begin{figure}
\centering
{\includegraphics[angle=-90,width=8cm]{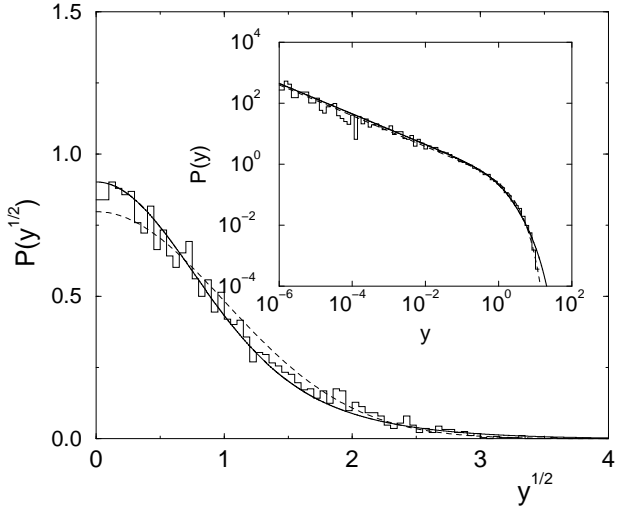}}
\caption{
Same as figures~\protect\ref{stat-m3} and \protect\ref{stat-m0}, 
but for the $m=2$ series.
The effect of scarring is
small and only marginal deviations from the
Porter-Thomas (Random Matrix Theory) distribution (dashed line)
are observed. 
}
\label{stat-m2}
\end{figure}

The same approach can be used for the data in other $m$ series. However, as is obvious in
Fig.~\ref{ft}(b), the contribution of the $z$ orbit is much smaller in e.g. the $m=3$ series.
As mentioned above, this is well understood semiclassically. In simple words, as the centrifugal term
is more important, it keeps the electron away from the $z$ axis and strongly diminishes the contribution of
this orbit. The $R$ parameter for the $m=3$ series can be extracted from the Fourier transform in Fig.~\ref{ft}(b)
and is close to 0.1. The semiclassical prediction, which can be calculated in the spirit of
refs.~\cite{Bogomolny89} and \cite{shaw95}, is $[R(m=0)]^4\approx 0.12$ in 
reasonably good agreement.
For such a small $R$ value, the deviation of the distribution~(\ref{modified_pt}) from Porter-Thomas
is very small. This explains why the Porter-Thomas distribution correctly reproduces
the results of the numerical experiment, see Fig.~\ref{stat-m3}.
We have also obtained results for the $m=1$ and $m=2$ series, shown
in Figs.~\ref{stat-m1} and \ref{stat-m2}. Significant deviations from
Porter-Thomas are observed, although smaller than for the $m=0$ series. Again, the
modified distribution, eq.~(\ref{modified_pt}), agrees very well with the numerical results.

An alternative approach to the statistical properties of the ionization widths is possible.
From the semiclassical approach, we know both the average trend and the modulations;
we can thus subtract (or rather divide) these factors out in order to concentrate on the fluctuations.
We thus rescale the numerical data to the expected average+oscillatory behavior,
that is define:
\begin{equation}
z_n = \frac{y_n}{\bar{y}(\hbar_n)}
\label{z}
\end{equation}
with $\bar{y}$ is defined in eq.~(\ref{ysc}).

\begin{figure}
\centering
{\includegraphics[angle=-90,width=8cm]{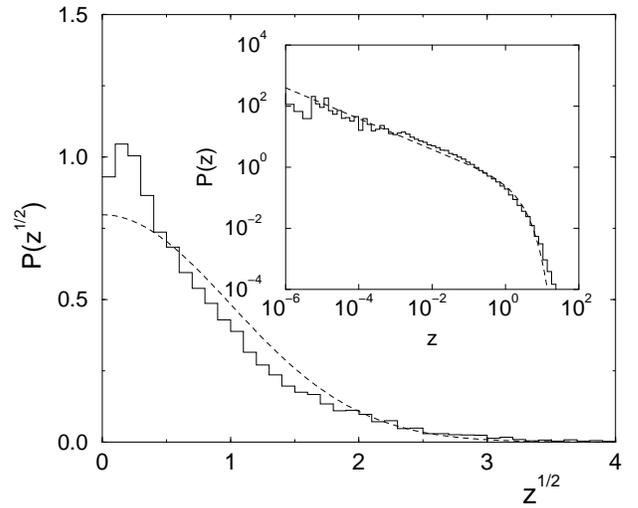}}
\caption{\label{pz-m0}
Statistical distribution of the ionization widths, rescaled according
to eq.~(\protect\ref{z}), to take into
account tunneling and the effect of the periodic orbit along the field axis.
A good agreement is obtained with the Porter-Thomas distribution, eq.~(\protect\ref{pt}),
shown as a dashed line. The data are the same as in Fig.~\protect\ref{stat-m0}.
The fact that a good agreement is obtained indicates that our model
describes scarring and tunneling in a satisfactory manner. 
}
\end{figure}

The statistical distribution of the $z$ variable is shown in Fig.~\ref{pz-m0}, for the $m=0$ series.
As can be seen, it agrees very well with a pure Porter-Thomas distribution. This fully confirms that,
once the average and oscillatory behavior have been properly taken into account, 
only the standard fluctuations described by Random Matrix Theory persist.

Finally, we have studied a slightly less realistic system: the two-dimensional hydrogen atom
in parallel electric and magnetic fields, obtained from the previous system by imposing
that the motion takes place in the $(x,z)$ plane. The classical dynamics is {\it exactly}
the same than for $m=0$ states (obviously the motion is planar for such cases).
One could thus naively expect the same properties for the ionization widths for the
quantum system.
This is however not entirely true for two reasons:
\begin{itemize}
\item The zero-point transverse motion is now in one dimension instead
of two. Thus, the shift energy is reduced by a factor 2 compared to eq.~(\ref{shift}):
\begin{equation}
\epsilon_{2D}  = \frac{1}{4} \gamma^{1/3} = \frac{1}{4} \hbar_{\mathrm{eff}}.
\end{equation}
The modified scaled energy, eq.~(\ref{hat_epsilon}), must be modified accordingly. 

\item The stability matrix is a $2\times 2$ matrix instead of a $4 \times 4$ matrix.
As explained in the appendix, this results in a denominators in eqs.~(\ref{cw1})-(\ref{cw4}),
to be square roots of the three-dimensional results for $m=0.$ 

\end{itemize} 

\begin{figure}
\centering
{\includegraphics[angle=-90,width=8cm]{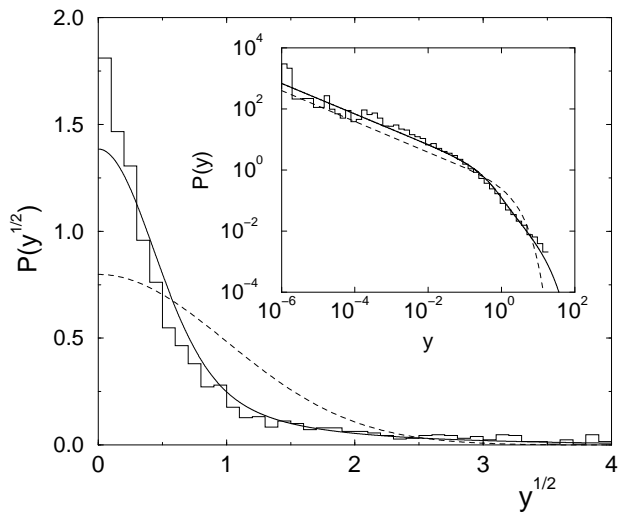}}
\caption{\label{stat-2d}
Same as figures~\protect\ref{stat-m3} and \protect\ref{stat-m0}, 
but for the two-dimensional hydrogen atom in parallel electric and magnetic fields.
Because of the reduced dimensionality, the motion transverse to the
$z$ periodic orbit is less unstable than in the three-dimensional 
atom, and the effect of scarring is
enhanced. Very large deviations from the
Porter-Thomas (Random Matrix Theory) distribution (dashed line)
are observed, but the improved model, eq.~(\protect\ref{modified_pt})
(solid line),
reproduces well the numerical results. 
}
\end{figure}

\begin{figure}
\centering
{\includegraphics[angle=-90,width=8cm]{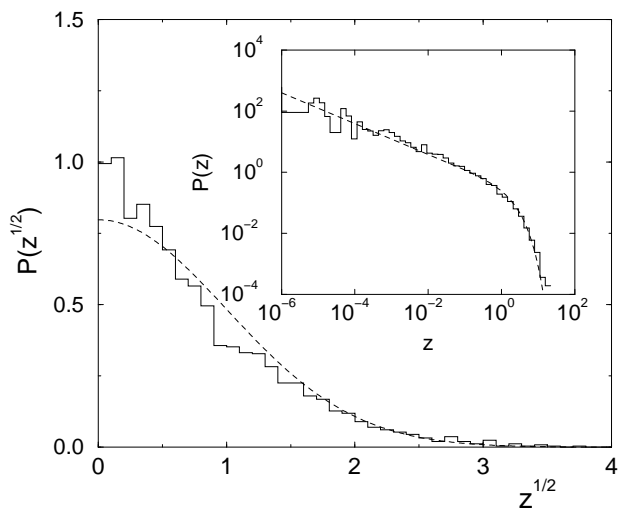}}
\caption{\label{pz-2d}
Same as Fig.~\protect\ref{pz-m0}, for the two-dimensional model of the
atom in parallel electric and magnetic fields (data as in Fig.~\protect\ref{stat-2d}).
Again, scarring is well accounted for by our model.
}
\end{figure}

The net effect is that the instability of the real orbit in the potential well
is significantly reduced, simply because there is less space for the electron
to escape far from the $z$ axis. The analysis is similar to the three-dimensional 
$m=0$ case, with the parameter $R$ being now taken at power 1/2, i.e. $R\approx 0.77$
instead of $R\approx 0.59.$ Stronger deviations from the Porter-Thomas
distribution are thus expected for the $y_n.$ Figs.~\ref{stat-2d} and \ref{pz-2d} show that it is indeed the
case. Once more, the agreement with the modified Porter-Thomas distribution,
eq.~(\ref{modified_pt}) is very good.

\section{Conclusion}

In this paper, we have studied the widths (ionization rates) of resonances
of a realistic system, the hydrogen atom in parallel electric and magnetic fields,
in conditions where the classical dynamics is chaotic. We have shown that, using
a semiclassical approach without any adjustable parameter but only with
classical ingredients, we are able to predict analytically the
average behavior of the widths. We have also shown the existence of a modulation of the
average width associated with a periodic orbit and calculated quantitatively 
its properties, again using only classical ingredients. Finally, the
residual fluctuations have been shown to be accurately described by a Random Matrix Model.
This proves that a proper combination of semiclassics and Random Matrix Theory 
can {\it predict} the behavior of the system vs. ionization. 

Our results are comparable to the ones obtained on a model system in~\cite{creagh02}.
For example, their Fig.~1 is clearly comparable to our Fig.~\ref{fp}. Note however that,
due to the specificities of our {\it realistic} system, the expressions
we obtain have a simpler form. On a different model system, Bies et al.~\cite{bies01}
observed also deviations from the Porter-Thomas distribution. Part of the
deviation is due to the relatively small value of the effective Planck constant,
but another part is certainly due to scarring. Their Figs.~4 and 5 are again very comparable
to our Fig.~\ref{fp}. Because they do not consider a scaling system, the classical
dynamics -- and consequently the properties of the periodic orbits -- 
change with energy which makes a comparison with our distribution rather difficult. 
We however have little doubt that the basic process at work is similar to ours.

\section{Acknowledgments}
We are grateful to Niall Whelan and Stephen Creagh for an initial stimulation
to look at tunneling in parallel fields problem. We thank W.E. Bies, L. Kaplan
and E.J. Heller for permission to use their data for our manipulations.
Support of KBN under project 5P03B-08821 (JZ) is acknowledged.
The additional support of the bilateral Polonium and PICS programs is
appreciated.
Laboratoire Kastler Brossel de
l'Universit\'e Pierre
et Marie Curie et de l'Ecole Normale Sup\'erieure is
UMR 8552 du CNRS. CPU time on various computers has been provided by IDRIS.

\appendix*

\section{Classical dynamics near the saddle point}

In this appendix, we discuss how the various classical quantities
which enter the semiclassical formula can be calculated in our specific
system, the hydrogen atom in parallel electric and magnetic fields.

The Hamiltonian of the system is given, in scaled units, by eq.~(\ref{hams}).
The saddle point is located along the $z$ axis at position:
\begin{equation}
z_{\mathrm{saddle}}= \frac{1}{\sqrt{f}},
\end{equation}
with energy $\epsilon_{\mathrm{ion}} = -2\sqrt{f}.$

As we are interested in highly excited states lying in the immediate
vicinity of the saddle point energy, it is convenient to expand the
Hamiltonian at second order around the saddle point. The normal
modes of this harmonic approximation are along the
$z$ axis and in the $x-y$ plane. In the $x-y$ plane, 
the saddle point is a potential minimum associated
with a vibration frequency:
\begin{equation}
\omega_{\rho} = \frac{\sqrt{1+4f^{3/2}}}{2}.
\label{omegarho}
\end{equation}
Because of the azimuthal symmetry around the fields axis, this mode is 
degenerate.
In order to have a chaotic motion in the inner potential well, the
scaled energy must be large, typically of the order of -0.1, which in turns
implies that $f$ is rather small. In most cases, one can thus forget the $f$ 
dependence
in eq.~(\ref{omegarho}) and use the approximation:
\begin{equation}
\omega_{\rho} \approx \frac{1}{2}.
\label{omegarhoapprox}
\end{equation}

Along the $z$ axis, the saddle point is a potential {\it maximum}. It is thus
associated with an eigenmode with purely imaginary frequency $i\omega_z$ where:
\begin{equation}
\omega_z=\sqrt{2} f^{3/4}
\label{omegaz}
\end{equation}
The corresponding imaginary period is nothing, but the period of the instanton.
Alternatively, $\omega_z$ can be viewed as the vibration frequency around the
saddle point in the {\it inverted} potential. In an harmonic potential, the
action of an orbit is simply (within a $2\pi$ factor) the ratio of its excitation energy
(with respect to the equilibrium point) to the frequency. This yields
the (imaginary) action of the instanton given by eq.~(\ref{K}).

The harmonic approximation around the saddle point can also be used for the
calculation of the stability matrix of the instanton. Indeed, as the
harmonic potential separates completely in a transverse and a longitudinal
component, the monodromy matrix of the instanton in each transverse direction,
after propagation during time $t,$
is simply of the form:
\begin{equation}
\begin{pmatrix}
\cos \omega_{\rho}t & -\sin \omega_{\rho}t  \\
\sin \omega_{\rho}t & \cos \omega_{\rho}t  
\end{pmatrix}
\label{mono}
\end{equation}
The stability matrix of the instanton is obtained by evaluating the monodromy
matrix at the period of the instanton $t=2i\pi/\omega_z:$
\begin{equation}
W=
\begin{pmatrix}
\cosh \frac{2\pi\omega_{\rho}}{\omega_z} & -i \sinh \frac{2\pi\omega_{\rho}}{\omega_z}  \\
i \sinh \frac{2\pi\omega_{\rho}}{\omega_z}  &  \cosh \frac{2\pi\omega_{\rho}}{\omega_z}
\end{pmatrix}
\label{w}
\end{equation}

In our case, the ratio $\omega_{\rho}/\omega_z$ is very large, so that the hyperbolic
trigonometric functions can be approximated by an exponential, yielding:
\begin{equation}
\sqrt{-\mathrm{det} (W-I)} \approx \exp{\left(\frac{\pi\omega_{\rho}}{\omega_z}\right)}
\label{det}
\end{equation}

For the three-dimensional hydrogen atom, the stability matrix is a $4\times 4$ matrix
which actually splits in two $2\times 2$ identical blocks (along the $x$ and $y$ directions)
of type~(\ref{w}). 
Thus, the contribution (\ref{det}) must be squared to get the correct
semiclassical contribution. In contrast, for the simplified
two-dimensional model, there is only one such contribution.

If one uses the approximate value (\ref{omegarhoapprox}) in eq.~(\ref{det}), one
finally gets the contribution (\ref{denum}).

The last ingredient in the semiclassical approximation is the 
stability matrix of the real periodic orbit in the inner potential well.
As explained in the main text, the series of bifurcations taking place in the
vicinity of the saddle point energy implies that this matrix changes rapidly
with $\epsilon.$ On the other hand, when $\epsilon$ is varied, the dynamics
inside the potential well is only weakly affected: the main effect is that the
electron spends less or more time in the immediate vicinity
on the saddle point. As the transverse potential is there rather steep,
the stability matrix varies a lot. These modifications are
essentially described by a multiplication by a matrix similar to (\ref{mono}).
A small variation of the period of the orbit is enough to affect strongly
the matrix. However, it is the product of the stability matrices of the
instanton and the real periodic orbit which describes the semiclassical
contribution, eq.~(\ref{cw4}). As it is the very same matrix type~(\ref{mono})
which contributes to the two stability matrix, it turns out that the
{\it modulus} of  $\det(WM-I)$ actually depends weakly on $\epsilon$ as shown
in Fig.~\ref{trace}.

\end{document}